\documentclass[reprint,amsmath,amssymb,aps,floatfix, pre]{revtex4-1}
\usepackage{verbatim}
\usepackage{graphicx}
\usepackage{caption}
\usepackage{dcolumn}
\usepackage{bm}
\usepackage{float}
\usepackage{wrapfig}
\usepackage{placeins}
\usepackage{ifthen}
\usepackage{xr}
\newcommand{\hext}{\ensuremath{h_\mathrm{ext}}}
\begin{document}
\title{Weird scaling for 2-D avalanches: \\Curing the faceting, and scaling in the lower critical dimension}
\author{L. X. Hayden}
\affiliation{ LASSP, Physics Department, Cornell University \\
 Ithaca, NY 14853-2501, United States}
 \author {Archishman Raju}
 \affiliation {The Rockefeller University, New York, NY 10065}
 \author{James P. Sethna}
 \affiliation{LASSP, Physics Department, Cornell University \\
 Ithaca, NY 14853-2501, United States}
\date{\today}
\begin{abstract}
The non-equilibrium random-field Ising model is well studied, yet there are outstanding questions. In two dimensions, power law scaling approaches fail and the critical disorder is difficult to pin down. Additionally, the presence of faceting on the square lattice creates avalanches that are lattice dependent at small scales. We propose two methods which we find solve these issues. First, we perform large scale simulations on a Voronoi lattice to mitigate the effects of faceting. Secondly, the invariant arguments of the universal scaling functions necessary to perform scaling collapses can be directly determined using our recent normal form theory of the Renormalization Group. This method has proven useful in cleanly capturing the complex behavior which occurs in both the lower and upper critical dimensions of systems and here captures the 2D NE-RFIM behavior well. The obtained scaling collapses span over a range of a factor of ten in the disorder and a factor of $10^4$ in avalanche cutoff. They are consistent with a critical disorder at zero and with a lower critical dimension for the model equal to two. 
\end{abstract}
\maketitle

We study the avalanche size distribution in the two-dimensional
nucleated non-equilibrium random-field Ising model (NE-RFIM), simulated on
a Voronoi lattice to bypass faceting, and analyzed using the scaling
predictions of the nonlinear renormalization-group flows predicted
for the lower critical dimension. We find excellent agreement over a
large critical region, addressing several outstanding issues in the field.

The NE-RFIM is perhaps the
best-understood model of crackling noise~\cite{SethnaDM01}, exhibiting
power-law distributions of avalanche sizes at a critical disorder $r_c$
representing the standard deviation of the strength of the random field
at each site. The model transitions from a `down-spin' state to an
`up-spin' state as an external field $H$ increases. Above the
critical disorder $r_c$, this transition is composed of avalanches of
spins of size limited by a typical cutoff $\Sigma_+(r)$; below the critical
disorder a finite fraction of the spins flip in a single event, with
precursors and aftershock sizes limited by $\Sigma_-(r)$. 
This model, albeit simple, contains the
necessary ingredients to describe hysteretic and avalanche behaviors in
a diverse set of systems. Barkhausen noise in magnets~\cite{Bertotti98}
decision making in socio-economics~\cite{Bouchaud13}, absorption and
desorption in superfluids~\cite{Lilly96, Detcheverry04} as well as the
effects of nematicity in high $T_c$ superconductors~\cite{Bonetti04,
Carlson06, Phillabaum2012} can each be understood in terms of `crackling
noise' naturally described by the NE-RFIM.

Although the NE-RFIM itself has been around in various forms since the
1970s~\cite{ImryMa75}, there are still a number of current questions 
and issues:

$\bullet$
{\em Is it in the same universality class as the equilibrium RFIM
model~\cite{BalogTarjusTissier18}?}
It has long been debated whether the equilibrium and non-equilibrium
versions of the model are in the same universality class. This question
of universality has been approached in a number of ways which have
suggested the same class for the two models~\cite{Maritan94,%
Perez-Reche04,Colaiori04,LiuDahmen09,LiuDahmen09-2,BalogTissierTarjus14}. 
Recent work using the non-perturbative RG indicate that the two models
are in different universality classes in lower dimensions~\cite{BalogTarjusTissier18}. Our
findings pretty clearly imply they are also different in two dimensions.

$\bullet$
{\em Is the lower critical dimension (LCD) two, 
or is power law scaling sufficient to capture the behavior in $D=2$?}
The equilibrium RFIM has been shown to have a LCD equal to 
two~\cite{BrayMoore85}, and the same is believed to be true for the
front-propagation variant of the NE-RFIM~\cite{Drossel98}.
For the nucleated model we study here, some
suggest that the LCD is two~\cite{Perkovic95,Perkovic96}, others
suggest that power-laws are
indeed able to capture the behavior and no crossover occurs in
2D~\cite{Spasojevic11, Spasojevic11-2}, and some suggest that a lower critical
dimension does not exist for this
model~\cite{Thongjaomayum13,Kurbah15,Shukla16,Shukla17}. Here, we 
derive the expected non-power-law scaling in the LCD from a nonlinear
renormalization-group analysis, and find excellent agreement with 
the data presuming an LCD of two, while power-law scaling fails to 
capture the behavior.

$\bullet$
{\em Is the value of the critical disorder in $D=2$ zero, or positive?}
In the nucleated model, the critical disorder
appears to decrease with dimension, going from $5.96 \pm 0.02$ in 5D to
$2.16 \pm 0.03$ in 3D~\cite{Sethna06}. This behavior in conjunction with
the observation that for both the equilibrium and front-propagation
problems, $r_c$ is found to be zero~\cite{Drossel98} suggests that $r_c$
may be quite small. Early work on the nucleated model, presuming
power law scaling~\cite{Vives95,Perkovic96,KuntzPhD},
yielded positive $r_c = 0.75\pm 0.03$~\cite{Vives95},
but more recent work on larger systems finds a smaller
$r_c=0.54\pm0.02$~\cite{Spasojevic11,Spasojevic11-2} 
collapsing over a small range $r \in [0.64, 0.70]$. Our non-power-law
scaling form would predict that power-law fits at a given system
size should succeed in small ranges of disorder, but that larger
system sizes will yield lower and lower predicted critical disorders.
Our results are compatible with a critical disorder of zero, 
directly (random field strength $r_c=0$) or perhaps more naturally
in conjunction with some random bond disorder (so $r_c<0$, see Appendix~\ref{app:Disorder}).

%Power law scaling collapses have long been a preferred method for
%demonstrating that the behavior of a critical system is well understood.
%That this type of heuristic procedure can work so well in such a
%widespread number of applications is initially surprising and leads
%naturally to the question of when and why this approach fails. For
%example, in the two-dimensional non-equilibrium random-field Ising model
%(2D NE-RFIM),  attempted collapses assuming power law scaling perform in
%very limited ranges of disorder~\cite{Spasojevic11, Spasojevic11-2,
%Perkovic96, KuntzPhD}, which we argue is a symptom of non-power law
%scaling. This failure can also be observed in a number of other systems,
%particularly at their lower and upper critical dimension. 

Scaling collapses ({\em e.g.}~Figs~\ref{fig:As_collapse} 
and~\ref{fig:dMdh_collapse}) are the gold standard for identifying
universal scaling behavior at critical points. Commonly used in 
simulations and experiments, the scaling form for a function
of two variables usually becomes a power law times a universal function of
the ratio of two power laws -- a result which follows from linearizing
the renormalization-group (RG) flows. The LCD, however, is precisely the
dimension at which one of the eigenvalues of the RG flow vanishes and the
nonlinear terms become crucial to the behavior. Recently, Raju 
{\em et al.~\cite{Raju17}} analyzed non-linearities in
renormalization group flows 
using normal form theory drawn from the dynamical systems community.
In the cases for which power laws work well, the dynamics are governed by a 
hyperbolic fixed point which can be linearized by a change of variables,
leading to traditional scaling predictions. Our simulations indicate
that the LCD for the NE-RFIM is poised at a transcritical bifurcation in
the RG flow. By considering the form the flow
equations should take, we are able to provide concrete non-power-law
invariant scaling variables which enable collapse of our data over a
range of a factor of ten in the disorder. This success, and the 
enormous critical region, suggests that using the appropriate invariant
scaling variables can be effective for analyzing experiments and simulations
systems at their LCD (like the XY model), despite exponentially growing
correlation lengths. (Similar analyses have been done for the 4-state Potts
model~\cite{SalasS97} and the XY model~\cite{PelissettoV13}, except
that their invariant scaling variables include only their predicted leading
log corrections.)

In addition to the application of our normal form theory of the
Renomalization Group, another key component to the success of our
collapses is an approach to dealing with the faceting. Running
simulations on a square lattice leads to distortions in the shape of the
distributions of interest due to lattice effects as the critical point
is approached. Long, unnaturally straight avalanche boundaries for small
disorder arise which serve to effectively decrease the simulation
size (Appendix~\ref{app:Faceting}).
To combat this, we run our simulations on a Voronoi
lattice. Although this introduces some intrinsic
disorder (Appendix~\ref{app:Disorder}), we find the Voronoi lattice to be
effective in combating faceting effects, enabling clean collapses over a
range of a factor of ten in the disorder, a significantly larger range
than the current available collapses which use data in a range $\approx
10\%$.

The paper is organized as follows. In Section~\ref{modeldefinition}, we define the model we will use in detail. In Section~\ref{normalform}, we work out the normal form of the RG flows. In Section~\ref{sec:invariant}, we derive the invariant scaling combinations used for the scaling collapses. We discuss the efficacy of our scaling collapses, our ability to fit parameters and alternative choices of scaling forms in Section~\ref{sec:parameter} before concluding.

\section{Model definition}
\label{modeldefinition}

\begin{figure}[!b]
\centering
  	\includegraphics[scale=0.08]{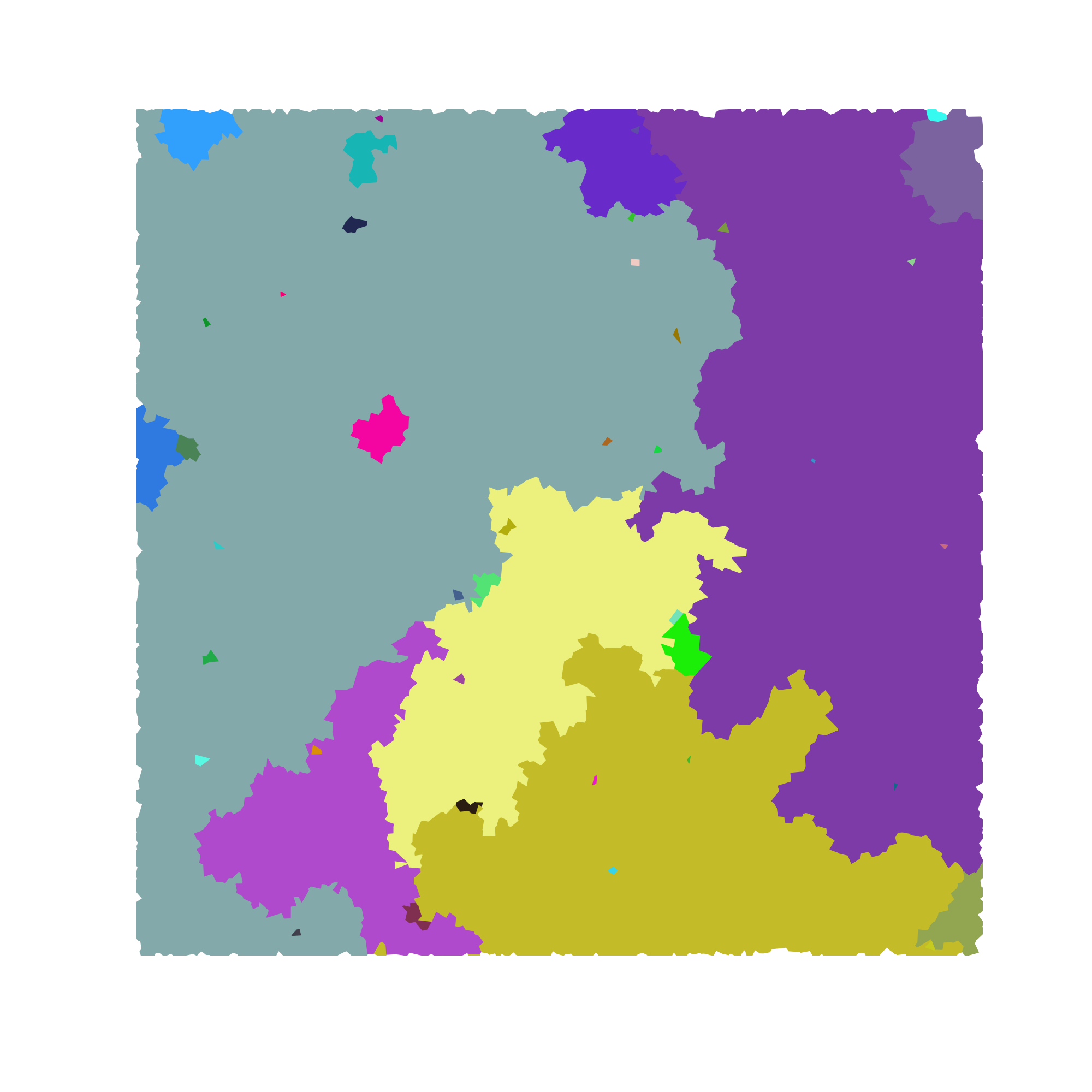}
  	\includegraphics[scale=0.08]{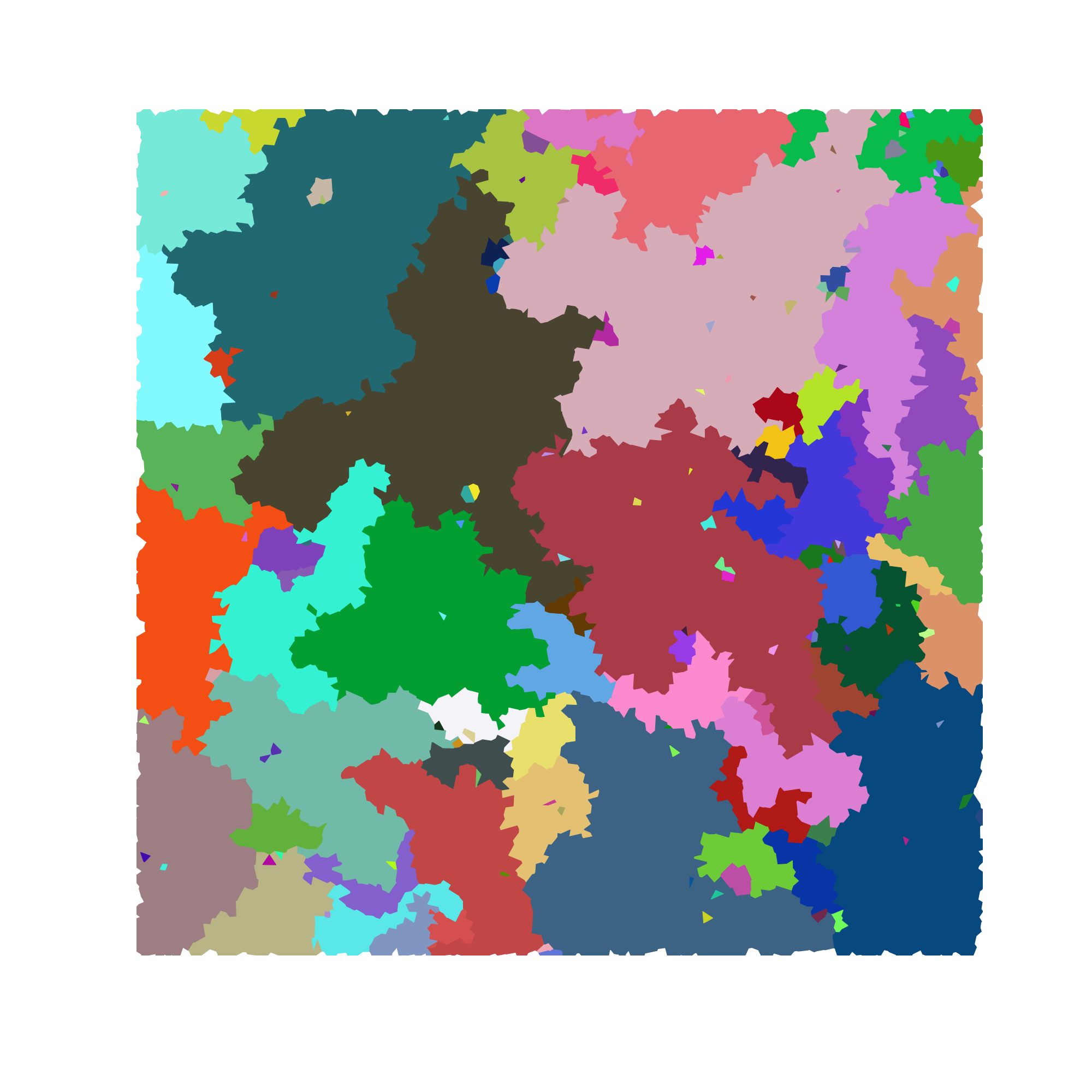}
  	\break
  	\includegraphics[scale=0.08]{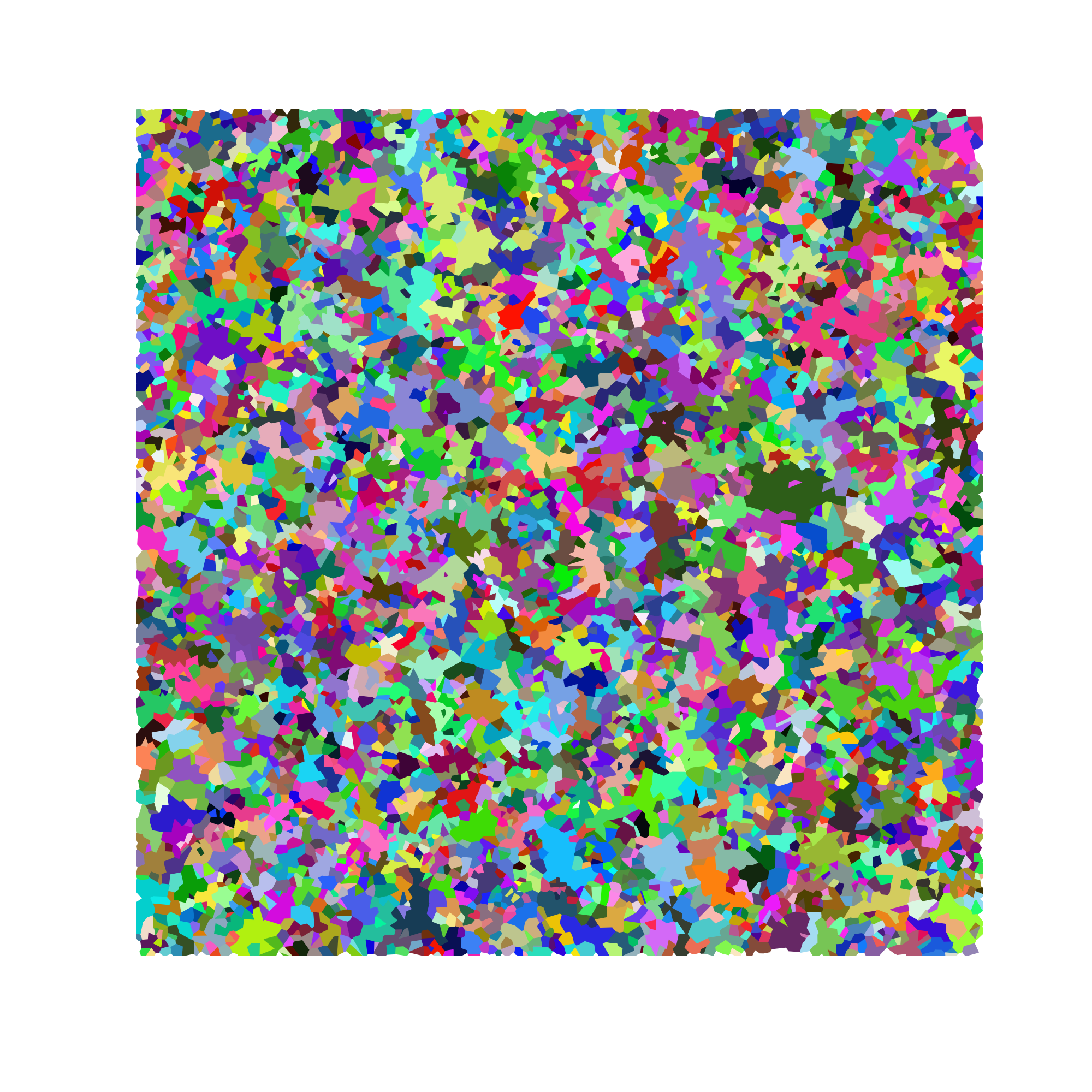}
  	\includegraphics[scale=0.08]{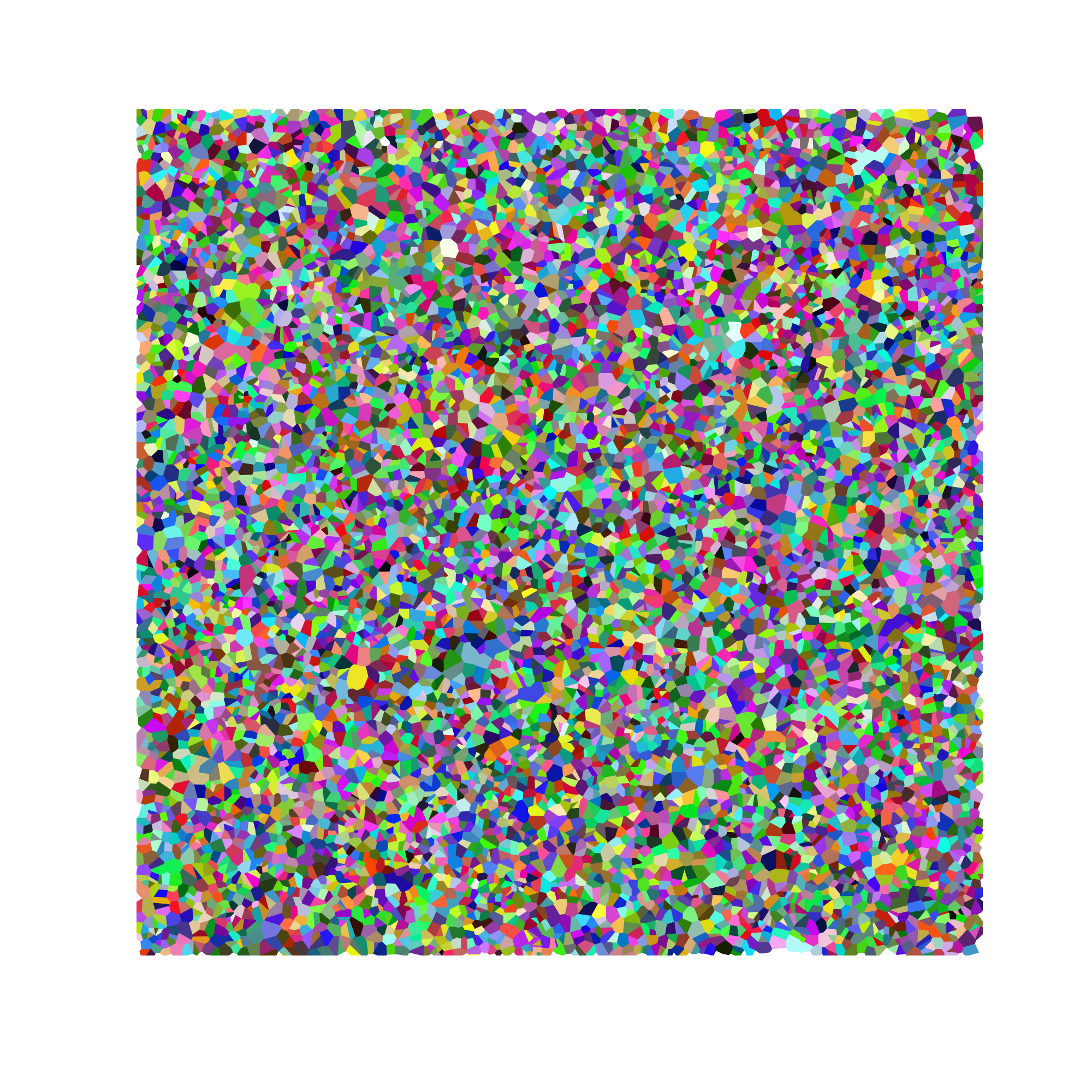}
  	\captionof{figure}{$r = 0.5,1.0,5.0, 50.0$ from left to right, top to bottom}
  	\label{fig:avalanches}
\end{figure}

The non-equilibrium random-field Ising model (NE-RFIM) consists of
Ising spins $S_i=\pm 1$
connected by bonds of strength $J=1$, subject both to a random field $h_i$
and an external field $\hext(t)$:
\begin{equation}
{\mathcal{H}} = -\sum_{\langle ij\rangle} J S_i S_j - \sum_i S_i (\hext + h_i).
\end{equation}
The external field starts at $\hext(t) = -\infty$ and grows until all spins
have flipped from $S_i = -1$ to $+1$. 
The spins flip when they can decrease the energy $H$, triggered either by an
increase in the external field $\hext(t)$ (spawning a new avalanche), or by
being kicked by the spin flip of a neighbor (propagating an existing
avalanche). The random fields are chosen from a distribution $p(h_i)$
with zero mean and a width $r$ describing the strength of the disorder
from the random field. The simulations in this work use the traditional
Gaussian form for the disorder,
$p(h_i) = (1/\sqrt{2 \pi r^2}) \exp(-h_i^2/2r^2)$.
At large disorder $r$ the avalanches are all small; as the disorder decreases
the avalanches grow in size (Fig.~\ref{fig:avalanches}). Avalanche size is denoted by $s$. There are 
two different variants of the NE-RFIM. The one we study is `nucleated' --
all spins start pointing down (-1), and the first avalanche is triggered
by a spin with an unusually large positive random field $h_i$. Another
model, inspired by fluid invasion into porous media, has a pre-existing 
front ({\em e.g.} a line of fluid-filled $+1$ spins at the bottom), and
does not allow for spins to flip unless at least one neighbor is up 
({\em i.e.}, with a path allowing fluid to enter). We study the nucleated
model, but occasionally refer to results from the front propagation model.

\begin{figure}
	\includegraphics[width=0.35\textwidth]{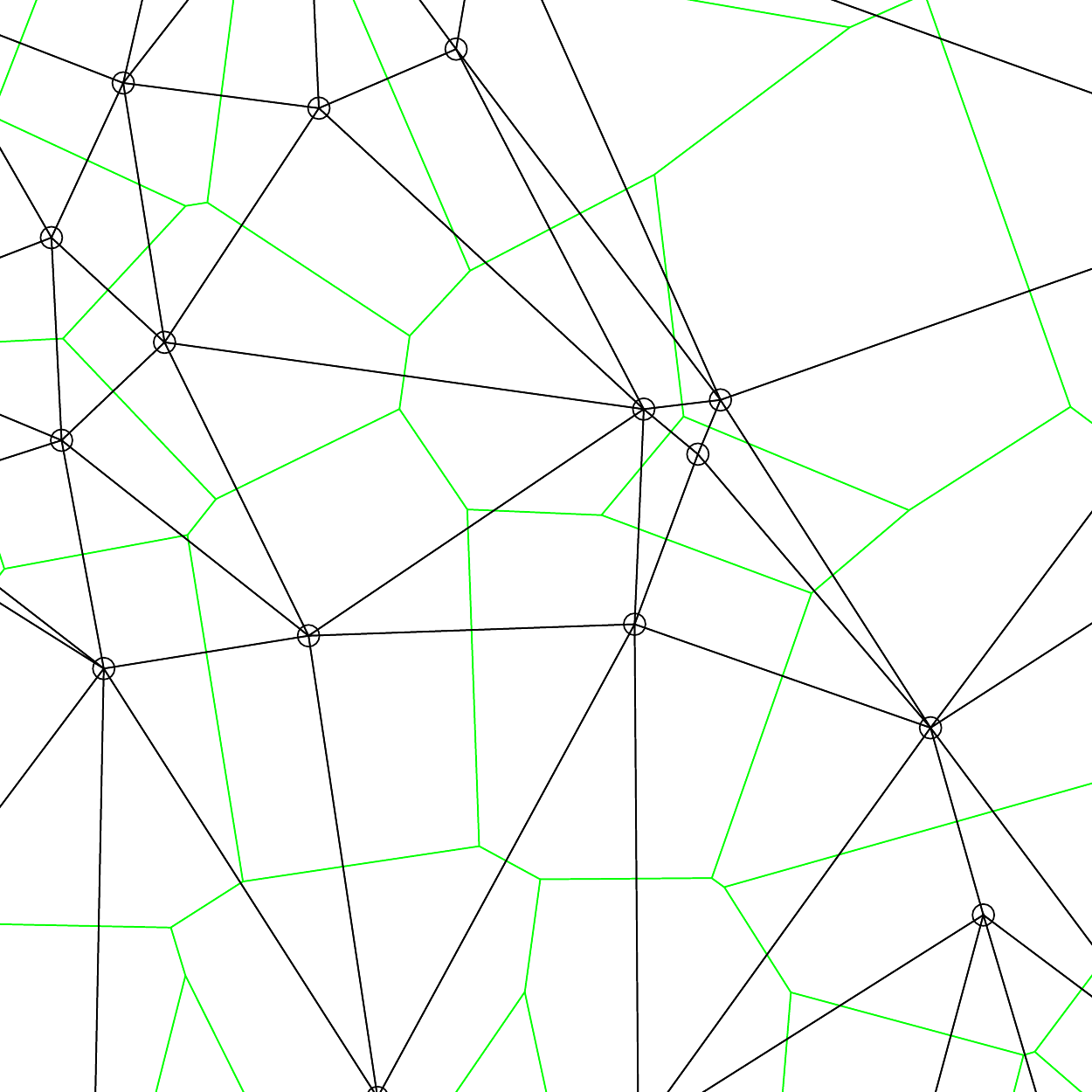}
	\caption{Segment of the Voronoi lattice. Spins at circled sites,
	randomly scattered in the two-dimensional plane. The spins interact
	with neighbors across bonds denoted by black lines. The Voronoi lattice
	in green determines the neighbors of each spin.}	
	\label{fig:Voronoi}
\end{figure}

Our simulations differ from tradition in that our spins are not on a regular
lattice. We work in two dimensions (where the behavior is still controversial),
but do not simulate spins on a square lattice but rather, as mentioned earlier, on a Voronoi
lattice (Fig.~\ref{fig:Voronoi}) -- with randomly scattered spins interacting
with nearest neighbors.
The neighbors are determined by shared boundaries
of Voronoi cells for each spin. The spin's Voronoi cell is the set of points
on the plane nearest to that spin.

\section{Renormalization Group Analysis}

\subsection{Flow equations}
\label{normalform}

Following the convention of Bray and Moore~\cite{BrayMoore85} for the
equilibrium model, we define a parameter $w$ which corresponds to the
ratio of the disorder $r$ over the coupling $J$ and determine its RG
flow equation through symmetry considerations. In principle, there are
an infinite series of terms. Using only analytic changes of variables,
however, it is possible to remove all terms of $O(4)$ or higher without
removing any universal behavior~\cite{Raju17}. We give a brief version of the argument here for completeness. 

In the equilibrium model, the flow equation is found to be $dw/d\ell=-(\epsilon/2)w+Aw^3+h.o.t.$ where $\epsilon=D-2$ and $w=r/J$~\cite{BrayMoore85}. For the NE-RFIM, however, $r$ has the symmetry $r \leftrightarrow -r$ while $J$ lacks this symmetry due to the external field. This implies $w \nleftrightarrow -w$ and suggests that the RG flow for $w$ in the NE-RFIM must include a squared order term. (Note that the symmetry $J \to -J$ for the equilibrium Hamiltonian is only
valid for systems with a bipartite lattice. It would be natural to test 
whether the equilibrium RFIM on a triangular lattice retains the pitchfork
form $\log \xi \sim 1/r^2$ or changes to the transcritical divergence
$\log \xi \sim 1/r$ as suggested by our symmetry argument.)

 Assuming the lower critical dimension $D=2$, we have $\epsilon=0$, and may choose a scale for the disorder $r_s$ such that the prefactor of the squared order term in the flow equation of $w$ is equal to one. Taking $J=1$, the choice we make for $w$ is $w=(r-r_c)/r_s$ where $r_c$ defines the critical disorder. The generic form for the flow equation of $w$ is given by
\begin{equation}
	\frac{dw}{d\ell}=w^2+B_1 w^3+ B_2 w^4 + \dots 
\end{equation}
\noindent Given such a flow equation with an infinite number of possible terms, normal form theory proceeds by systematically removing higher order terms with a change of variables. Consider the change of variables $w = \tilde{w} + b_1\tilde{w}^2+b_2\tilde{w}^3+b_3\tilde{w}^4+\dots$ The resulting flow equation takes the form:
\begin{equation}
	\frac{d\tilde{w}}{d\ell}=\tilde{w}^2+B_1 \tilde{w}^3+ (B_1 b_1+b_1^2+B_2-b_2) \tilde{w}^4 + \dots
\end{equation} 
\noindent With an appropriate choice of $b_1$ and $b_2$, the coefficient of $\tilde{w}^4$ may easily be set to zero. Likewise, all higher order terms may be systematically removed. Dropping the tildes and subscripts for clarity, the final form of the flow equation is given by 
\begin{equation}
	\label{eq:TruncNF}
	\frac{dw}{d\ell}=w^2+B w^3
\end{equation}

\noindent which corresponds to the normal form of a transcritical bifurcation~\footnote{The traditional transcritical bifurcation normal form~\cite{Strogatz14} $dw/d\ell = w^2$ is derived using the implicit function theorem, but involves changes of variables that alter critical properties in singular ways. Eq.~\ref{eq:TruncNF} is the simplest form that can be reached by successive polynomial changes of variables.}.

Next consider the flow equations for $s$ and $h$. The eigenvalues for these are given by $\lambda_s=d_f$ and $\lambda_h$ respectively where $d_f$ denotes the fractal dimension. In each case, the zero eigenvalue of $w$ gives rise to cross terms between $s$ and $w$ and $h$ and $w$. Again, in principle, we have an infinite number of possible terms but most all terms may be removed with a polynomial change of variables. The flow equations for $s$ and $h$ are hence given by
\begin{equation}
	\begin{split}
		&ds/d\ell= -d_f s-C s w ,\\ 
		&dh/d\ell= \lambda_h h+F h w
	\end{split}
\end{equation}
\noindent where in higher dimensions $d_f = 1/\sigma \nu$ and $\lambda_h = \beta \delta / \nu$. In two dimensions, the individual exponents $\sigma \to 0$ and
$\nu$ and $\beta \delta \to \infty$, keeping the combinations we use finite.  The coefficients $B$, $C$, and $F$ are {\em universal}. Just as the linear terms at ordinary (hyperbolic) fixed points yield universal critical exponents, these terms control universal dependences of physical behavior with changes in the control parameters. Note that, while they cannot be set to zero by a coordinate change, they may have universal values equal to zero~\footnote{For example, a term corresponding to $F h w$ in the flow equations of the magnetic field turns out to be equal to zero in the 4-d Ising model}.

\subsection{Invariant Scaling combinations}
\label{sec:invariant}

The appropriate scaling variables to collapse the data can be directly calculated from the flow equations (see Appendix~\ref{app:invariant} for full derivation).  We may directly solve for the correlation length $\xi\sim(1/w + B)^{-B}\exp(1/w)$ in the normal form variables by integrating Eq.~\ref{eq:TruncNF}.  The invariant scaling combination for $s$ obtained takes the form $s/\Sigma(w)$ where $\Sigma(w)$ is a nonlinear function of $w$. We allow for an undetermined scale factor $\Sigma_s$. The resulting form is given by 
\begin{equation}
 	\label{eq:SigmaOfw}
	\Sigma(w) = \Sigma_s (B+1/w)^{-B d_f+C}\exp(d_f/w).
\end{equation}
\noindent Likewise for $h$, we obtain:
\begin{equation}
 	\label{eq:etaOfw}
	\eta(w) = \eta_s(B+1/w)^{B \lambda_h-F}\exp(-\lambda_h/w),
\end{equation}
\noindent where $(h-h_{max})/\eta(w)$ is invariant under the RG, and $\eta_s$ is another scale factor. 

\begin{figure}
	\includegraphics[scale=0.3]{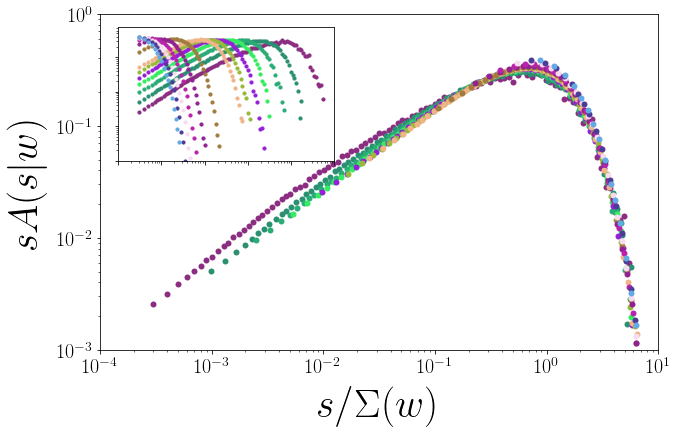}
  	\caption{Scaling collapse of the area weighted avalanche size distribution $A(s|w)$ for $w$ ranging from $0.8$ to $8.0$. There is a slight bulge at $s/\Sigma(w)\sim10^{-2}$ for small $w$. The shape of this curve $\mathcal{A}$
is universal -- it 
should be reproduced in experiments and other simulations in the same
universality class.}
  	\label{fig:As_collapse}
\end{figure}

First consider the area weighted size distribution $A(s|w)$. In analogy with three dimensions, we take $A(s|w) = s^{-1}v_s^x \mathcal{A}(v_s^y)$ where $v_s$ is the scaling variable and the prefactor of $s^{-1}$ arises from normalization constraints with $v_s=s/\Sigma(w)$ from Equation~\ref{eq:SigmaOfw}. The avalanche size distribution also depends on an unknown universal scaling function, $\mathcal{A}$. In order to perform our fits, we choose functional forms for the universal scaling functions. For the area weighted avalanche size distribution, we choose 
\begin{equation}
	\label{eqn:univA}
	\mathcal{A}(v_s) = \frac{1}{\mathcal{A}_N}v_s^{a_1}\exp(v_s^{a_2})
\end{equation} 
\noindent where the leading power law $v_s^x$ has been absorbed into $v_s^{a_1}$ here and $\mathcal{A}_N$ is the normalization factor $\mathcal{A}_N = \big{[}\Gamma\big{(}\frac{a_1}{a_2}\big{)}\gamma\big{(}\frac{a_1}{a_2}, \Sigma(w)^{-2a_2}\big{)}\big{]}/a_2$ where $\gamma$ denotes the regularized upper incomplete gamma function. The associated collapse is shown in Figure \ref{fig:As_collapse}. The best-fit values of the fitting parameters are $a_1 = 0.6955$ and
$a_2 = 1.1057$, which yield an approximation to the universal scaling
function $\mathcal{A}$
(Fig.~\ref{fig:As_collapse}).

\begin{figure}
	\includegraphics[scale=0.3]{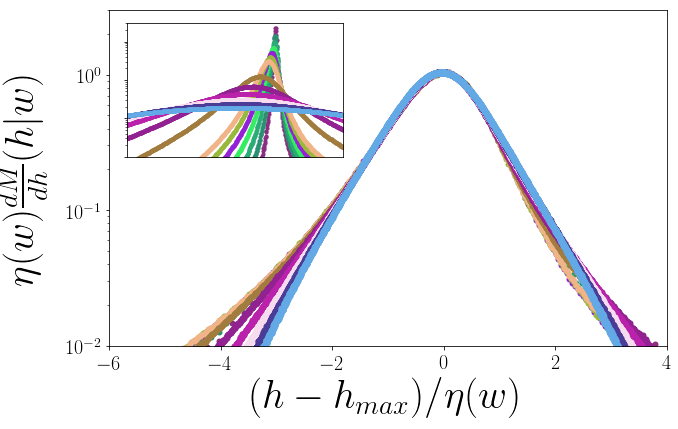}
  	\caption{ Scaling collapse of the change in magnetization of the sample with respect to the field $\frac{dM}{dh}(h|w)$ for values of $w$ ranging from $0.8$ to $8.0$. Again, the shape of this curve is universal.}
  	\label{fig:dMdh_collapse}
\end{figure}

Likewise, in analogy with three dimensions, we obtain $dM/dh(h|w)=\eta(w)^{-1} d\mathcal{M}/dh(v_h)$ where $v_h=(h-h_{max})/\eta(w)$ is the invariant scaling 
variable. For $d\mathcal{M}/dh$ we choose
\begin{equation}
	\frac{d\mathcal{M}}{dh}(v_h)=   \frac{1}{\frac{d\mathcal{M}}{dh}_{N}} \exp\bigg{[}\bigg{(}\frac{-v_h^2}{m_1+m_2 v_h+m_3 v_h^2}\bigg{)}^{m_4/2}\bigg{]}
\end{equation}
\noindent where $v_h = (h-h_\mathrm{max})/\eta(r)$, and 
$\frac{d\mathcal{M}}{dh}_{N}$ is a normalization factor computed as a 
sum of $\frac{dM}{dh}$ over the data range. The associated collapse is shown in Figure \ref{fig:dMdh_collapse}. The best-fit values of the fitting parameters are $m_1 = 0.5748$,
$m_2 = -0.1658$, $m_3 = 0.3563$, and $m_4 = 1.3449$, approximating our
prediction for the universal scaling function $\mathcal{M}$ 
(Fig.~\ref{fig:dMdh_collapse}).
 

\section{Parameter values}
\label{sec:parameter}

\begin{figure}
		\includegraphics[width=0.4\textwidth]{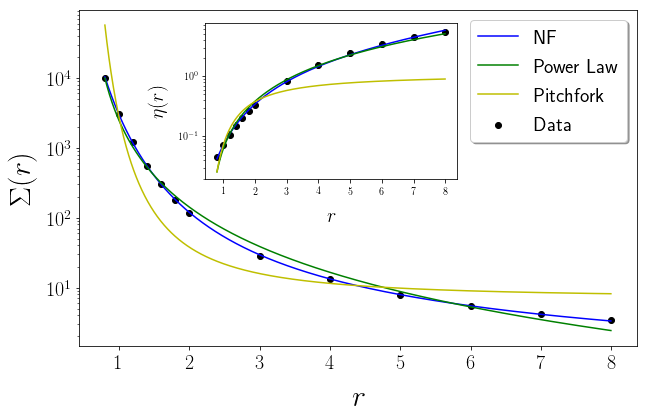}
		\caption{ Comparison of the best fit of $\Sigma(w)$ and $\eta(w)$ derived with different functional forms of $\frac{dw}{dl}$. We have $w=(r-r_c)/s_s$ such that $\Sigma(r)=\Sigma(w)$ and $\eta(r)=\eta(w)$. `NF' corresponds to $\Sigma$ and $\eta$ derived from the transcritical normal form, `Power Law' the hyperbolic (power law) form and `Pitchfork' the pitchfork form.}
		\label{fig:comparison}
\end{figure}

Through performing the scaling collapses we are provided with values of $\Sigma$ and $\eta$ for each value of disorder, $r$.  Using the nonlinear scaling forms for each of these we may then extract values for the associated parameters. An unconstrained fit yields a fractal dimension larger than two, the dimension of the system, which is unphysical. The 2D avalanches we consider appear compact. This suggests that the fractal dimension should be given by $d_f=2$ and that the maximum avalanche size should scale as the square of the correlation length. For this reason, we expect also that $\Sigma(w)\sim\xi^2$ and set $C=0$. Imposing these constraints, the fits obtained are able to describe the data well, as shown in Figure~\ref{fig:comparison}.

As usual, our data is precise enough that the statistical errors in the 
parameters we estimate are small compared to various systematic errors.
The dependence of our estimated $\Sigma(w)$ and $\eta(w)$ on the 
range of data and functional form appear smaller than the datapoints
in Fig.~\ref{fig:comparison}. 
We explore the importance of finite size effects and lattice effects
at small and large $r$ by performing the collapses and subsequent fits
of the nonlinear forms using subsets of the disorders for which we
have data [11 out of 13 points]. The best-fit parameters, with error
estimates given by the standard deviation of these measurements, are
given in the NF column of Table~\ref{tab:params}. Even larger
uncertainties, estimated in the last column, arise from excellent
fits that test various conjectures about the parameters.

Note that the best fit value of $r_c$ is found to be less than zero.
There are several possible explanations for this. One, $r_c<0$ could
indicate the Voronoi lattice used introduces an amount of intrinsic
disorder(Appendix~\ref{app:Disorder}).
This is certainly plausible as random bond and random field
disorder are expected to belong to the same universality
class~\cite{Dahmen96, Vives95}. Alternatively, constraining $r_c=0$ we
obtain a comparable fit by including an alternative normal form,
$NF_{\textrm{alt}}$, differing from $\Sigma(w)$ and by analytic
corrections to scaling (expected for the larger disorders
considered, see Appendix~\ref{app:well-behaved}). In either case, the results are
consistent with $r_c=0.$

\begin{table*}
	\centering
	\begin{tabular}{|c|c|c|c|c|c|}
	\hline
	& $NF$ & $NF_0$ & $NF_{\textrm{alt}}$ & $NF_{\textrm{Harris}}$ & Conjecture\\
	\hline
	$r_c$ & $-0.46 \pm 0.06$ & \textbf{0} &  \textbf{0} & $-0.46\pm0.06$ & $[-0.5,0.0]$\\
	$\lambda_h$ & $0.52 \pm 0.07$ & $0.24\pm0.08$ & $0.70\pm0.05$ & \textbf{1} & $1$\\
	$B$ & $-0.15 \pm 0.01$ & $0.039\pm0.007$ & $-0.76\pm0.14$ & $-0.25\pm0.03$ & $[-0.8, 0.0]$\\
	$F$ & $1.33 \pm 0.12$ & $2.02\pm0.13$ & $0.45\pm0.04$ & $0.45\pm0.06$ & $[0.0,0.5]$\\
	$C$ & \textbf{0} & $1.76\pm0.28$ &  \textbf{0} & \textbf{0} & 0\\
	$d_f$ & \textbf{2} & \textbf{2} & \textbf{2} & \textbf{2} & 2\\
	\hline
	\end{tabular}
	\caption{
	Table of the parameter values determined through a joint fit
	of $\Sigma(w)$ and $\eta(w)$. $NF$ corresponds to the
	transcritical form and $NF_{\textrm{alt}F}$ to the
	alternative transcritical form described
	in Appendix~\ref{app:well-behaved}. $NF_0$ corresponds to the transcritical
	form with $r_c=0$ and $NF_{\textrm{Harris}}$ to
	$\lambda_h=1$, the Harris criteria. To compute the error
	bars,  we performed the collapses and subsequent fits of the
	nonlinear forms using subsets of the disorders for which we
	have data [11 out of 13 points]. The errors given are the
	standard deviation of the values determined in this way.
	Values in bold were fixed in the corresponding fit.
	(Nonuniversal parameters in Tables~\ref{tab:wellbehavedcompare} and ~\ref{tab:truncatedcompare})
}
	\label{tab:params}
\end{table*}
As a test of our finding that the 2D NE-RFIM corresponds to a transcitical bifurcation, we may compare the fits obtained to those using different underlying assumptions. In particular, it is straightforward to calculate $\Sigma$ and $\eta$ assuming a hyperbolic fixed point (corresponding to power law scaling) and a pitchfork bifurcation (Appendix~\ref{app:pitchfork}). For each of these cases we can perform a fit to the values of $\Sigma(w)$ and $\eta(w)$ extracted from the collapse. The comparison of these fits are shown in Figure~\ref{fig:comparison}.

\begin{figure}
		\includegraphics[width=0.4\textwidth]{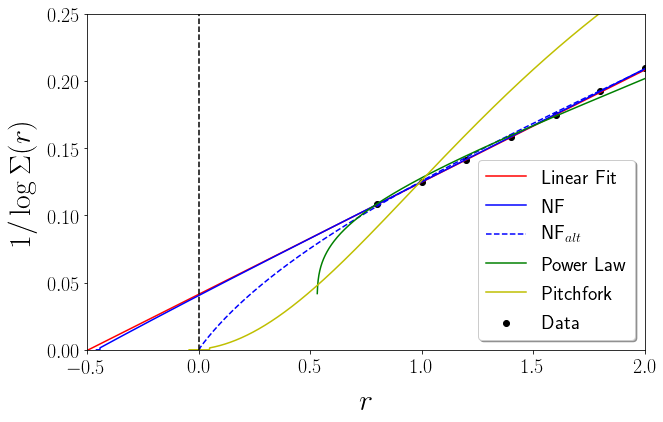}
		\caption{Comparison of $1/\log\Sigma(w)$ for the best fit of $\Sigma(w)$ derived with different functional forms of $\frac{dw}{dl}$. We have $w=(r-r_c)/s_s$ such that $\Sigma(r)=\Sigma(w)$. `NF' corresponds to $\Sigma$ derived from the transcritical normal form, `NF$_\mathrm{alt}$' to an alternative
normal form (Appendix~\ref{app:well-behaved}) constraining $r_c=0$,
`Power Law' the hyperbolic (power law) form and `Pitchfork' the pitchfork form.}
		\label{fig:logplot}
\end{figure}

It is particulary illuminating to consider the behavior of $1/\log\Sigma(w)$. For a transcritical bifurcation, the exponential divergence (ignoring $B$ and $C$ in Equation~\ref{eq:SigmaOfw}) gives $1/\log\Sigma(w) \sim w/d_f.$
Hence, if the behavior corresponds to a transcritical bifurcation, we would expect a plot of $1/\log\Sigma$ to scale linearly with the disorder. A comparison of the linear fit to $1/\log\Sigma$, along with the plots of $1/\log\Sigma$ for the best fits with a power law and pitchfork form are shown in Figure~\ref{fig:logplot}.
\noindent The results clearly support a transcritical bifurcation, perhaps
with $r_c<0$ (Appendix~\ref{app:Disorder}),
and challenge the alternative power law and pitchfork assumptions.

Simulation data of the 2D non-equilibrium random-field Ising model on a lattice which suppresses faceting is explained well by the presence of a transcritical bifurcation, and is incompatible with power law scaling or pitchfork normal forms without large corrections to scaling. This provides evidence that (1)~the universality class of the equilibrium and non-equilibrium models are indeed different and that (2)~power law scaling (which is governed by a hyperbolic fixed point) is not the correct approach for this system in this regime. The latter conclusion, in turn, is consistent with (3)~the LCD of the model being equal to two, or perhaps close to two.   

\begin{figure}[ht]
	\includegraphics[width=0.4\textwidth]{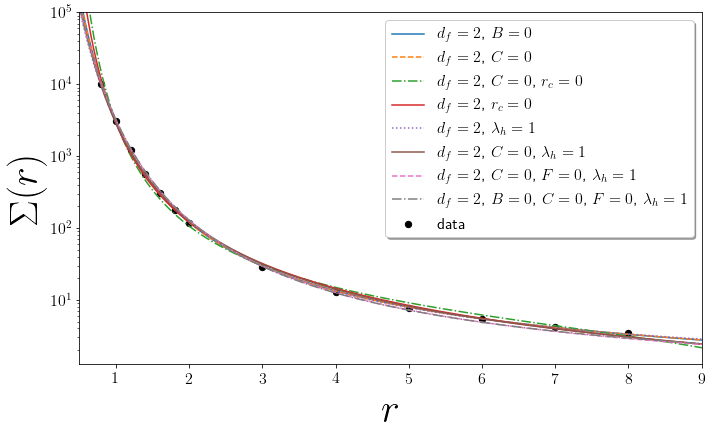}
	\caption{Fit comparisons $\Sigma_{\textrm{th}}(w)$, transcritical form}	
\label{fig:Sigmatruncated}
\end{figure}

\begin{figure}[ht]
	\includegraphics[width=0.4\textwidth]{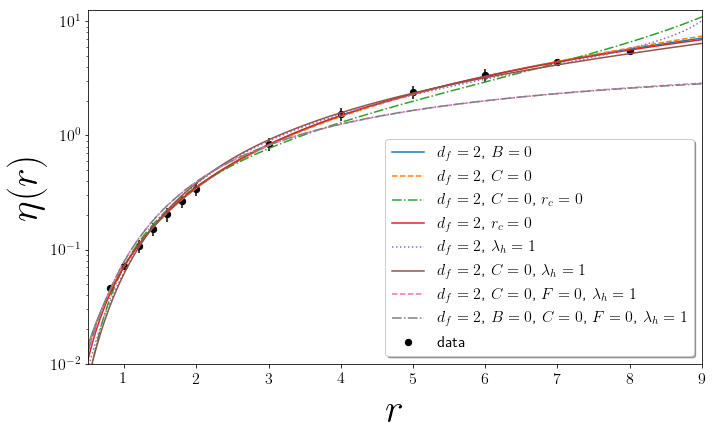}
	\caption{Fit comparisons $\eta_{\textrm{th}}(w)$, transcritical form}
	\label{fig:etatruncated}
\end{figure}

\begin{figure}[ht]
	\includegraphics[width=0.4\textwidth]{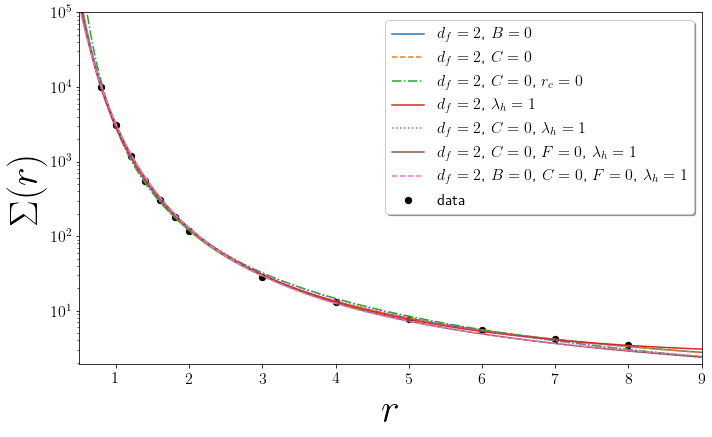}
	\caption{Fit comparisons $\Sigma_{\textrm{alt}}(w)$, alternative transcritical form}
	\label{fig:Sigmawellbehaved}
\end{figure}

\begin{figure}[ht]
	\includegraphics[width=0.4\textwidth]{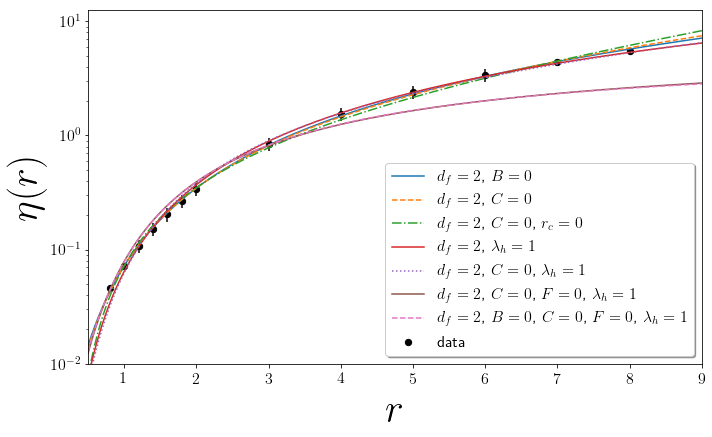}
	\caption{Fit comparisons $\eta_{\textrm{alt}}(w)$, alternative transcritical form}
	\label{fig:etawellbehaved}
\end{figure}

Although the transcitical bifurcation provides the best description of our simulation data, the corresponding parameter values are difficult to pin down. There are a number of restrictions we can make to the parameter values and still obtain a reasonable joint fit of $\Sigma(w)$ and $\eta(w)$ For example, we may require that the Harris criteria saturates, that $r_c=0$~\cite{Perkovic96} or that the coefficient of the quintic order term $B=0$. Each of these provides a good description of our data. A wide range of fits with various restrictions are shown in Figures~\ref{fig:Sigmatruncated}, ~\ref{fig:etatruncated}, ~\ref{fig:Sigmawellbehaved}, and~\ref{fig:etawellbehaved}. Corresponding best fit parameter values are shown in Tables~\ref{tab:truncatedcompare} and~\ref{tab:wellbehavedcompare}. As anticipated, the alternative form for the transcritical bifurcation is able to better capture the behavior far from the critical point.

%\break


\begin{table*}
	\begin{tabular}{|c|c|c|c|c|c|c|c|c|}
	\hline
	& & $NF$ & &  $NF_0$ & & $NF_{\textrm{Harris}}$  \\
		\hline
		$r_s$ & $5.11 \pm 0.54$ & $5.49 \pm 0.18$ & $2.91 \pm 0.27$ & $2.04 \pm 0.34$&  $6.89 \pm 0.63$ & $4.93 \pm 0.15$ & $6.78 \pm 0.20$ & $7.40 \pm 0.12$ \\
		$r_c$ & $-0.42 \pm 0.11$ & $-0.46 \pm 0.06$ & \textbf{0} & \textbf{0}&  $-0.65 \pm 0.10$ & $-0.46 \pm 0.06$ & $-0.64 \pm 0.06$ & $-0.70 \pm 0.03$ \\
		$\Sigma_s$ & $1.24 \pm 0.66$ & $1.11 \pm 0.05$ & $5.27 \pm 1.96$ & $14.40 \pm 8.84$&  $0.68 \pm 0.22$ & $1.64 \pm 0.04$ & $0.64 \pm 0.04$ & $0.54 \pm 0.005$ \\
		$\eta_s$ & $3.16 \pm 0.11$ & $3.11 \pm 0.44$ & $1.02 \pm 0.52$ & $0.50 \pm 0.41$&  $6.26 \pm 0.25$ & $4.33 \pm 0.42$ & $5.55 \pm 0.52$ & $6.08 \pm 0.82$ \\
		$d_f$ & \textbf{2} & \textbf{2} & \textbf{2} & \textbf{2}& \textbf{2} & \textbf{2} & \textbf{2} & \textbf{2}\\
		$\lambda_h$ & $0.44 \pm 0.04$ & $0.52 \pm 0.07$ & $0.61 \pm 0.05$ & $0.24 \pm 0.08$& \textbf{1} & \textbf{1} & \textbf{1} & \textbf{1} \\
		$B$ & \textbf{0} & $-0.15 \pm 0.01$ & $-0.27 \pm 0.03$ & $0.039 \pm 0.007$ & $ -0.69 \pm 0.21$ & $-0.25 \pm 0.03$ & $-0.09 \pm 0.01$ & \textbf{0} \\
		$C$ & $0.46 \pm 0.10$ & \textbf{0} & \textbf{0} & $1.76 \pm 0.28$ &  $-1.38 \pm 0.52$ & \textbf{0} & \textbf{0} & \textbf{0} \\
		$F$ & $1.72 \pm 0.08$ & $1.33 \pm 0.12$ & $0.73 \pm 0.02$ & $2.02 \pm 0.13$&  $-0.37 \pm 0.35$ & $0.45 \pm 0.06$ & \textbf{0} & \textbf{0} \\
		\hline
	\end{tabular}
	\caption{\label{tab:truncatedcompare} Table of the best fit values corresponding to Figures~\ref{fig:Sigmatruncated} and~\ref{fig:etatruncated}. Values in bold correspond to values fixed in the fit.}

	\begin{tabular}{|c|c|c|c|c|c|c|c|}
	\hline
	& & & $NF_{\textrm{alt}}$  \\
		\hline
		$r_s$ & $5.10 \pm 0.54$ & $5.05 \pm 0.36$ & $2.12 \pm 0.33$ & $1.81 \pm 0.08$ & $3.62 \pm 0.18$ & $6.57 \pm 0.39$ & $7.40 \pm 0.12$ \\
		$r_c$ & $-0.42 \pm 0.11$ & $-0.42 \pm 0.09$ & \textbf{0} & $ -0.15 \pm 0.07$ & $-0.29 \pm 0.09$ & $-0.62 \pm 0.10$ & $-0.70 \pm 0.03$ \\
		$\Sigma_s$ & $1.24 \pm 0.66$ & $-1.27 \pm 0.38$ & $13.16 \pm 5.72$ & $21.44 \pm 1.31$ & $3.19 \pm 0.45$ & $-0.67 \pm 0.32$ & $ 0.54 \pm 0.005$ \\
		$\eta_s$ & $3.16 \pm 0.11$ & $2.49 \pm 0.20$ & $0.56 \pm 0.40$ & $0.69 \pm 0.04$ & $2.48 \pm 0.09$ & $5.42 \pm 0.17$ & $6.08 \pm 0.82$ \\
		$d_f$ & \textbf{2} & \textbf{2} & \textbf{2} & \textbf{2}& \textbf{2} & \textbf{2} & \textbf{2} \\
		$\lambda_h$ & $0.44 \pm 0.04$ & $0.54 \pm 0.08$ & $0.70 \pm 0.05$ & \textbf{1} & \textbf{1} & \textbf{1} & \textbf{1} \\
		$B$ & \textbf{0} & $ -0.24 \pm 0.01$ & $-0.76 \pm 0.14$ & $-1.70 \pm 0.16$  & $-0.56 \pm 0.05$ & $-0.13 \pm 0.01$ & \textbf{0} \\
		$C$ & $0.46 \pm 0.10$ & \textbf{0} & \textbf{0} & $-0.31 \pm 0.04$ & \textbf{0} & \textbf{0} & \textbf{0} \\
		$F$ & $1.72 \pm 0.08$ & $1.22 \pm 0.16$ & $0.45 \pm 0.04$ & $-0.031 \pm 0.038$ & $0.35 \pm 0.03$ & \textbf{0} & \textbf{0}\\
		\hline
	\end{tabular}
	\caption{\label{tab:wellbehavedcompare} Table of the best fit values corresponding to Figures~\ref{fig:Sigmawellbehaved} and~\ref{fig:etawellbehaved}. Values in bold correspond to values fixed in the fit.}
\end{table*}

In three and higher dimensions~\cite{Perkovic96,Kuntz00}, measuring a
variety of avalanche properties was crucial in pinning down the
universal critical exponents and scaling functions. $dM/dH$ and the
cumulative avalanche size distribution, measured here, were supplemented by
measurements of finite-size scaling, avalanche correlation functions,
avalanche sizes binned in $H$, spanning avalanches, avalanche durations,
and average avalanche temporal shapes. Larger system sizes should be possible
with improved Voronoi data structures: the intercept of
Fig.~\ref{fig:logplot} suggests that a random-field free $r=0$ 
simulation of size $L \sim \sqrt(\Sigma(r=0)) = e^{10} \approx 22,000$
might divide into multiple avalanches at $r=0$, implying that $r_c<0$.

\section{Conclusions}
\label{conclusions}

In summation, performing large scale simulations on a Voronoi lattice and analyzing the RG flow equations yields valuable insight into the behavior of the NE-RFIM in 2D. The data collapses in a range of a factor of ten in the disorder and a factor of $10^4$ in the avalanche cutoff. The scaling is consistent with a critical disorder of zero and with a lower critical dimension of two. 

\begin{acknowledgments}
This work was partially supported by NSF grants  DMR-1719490 and DGE-1144153. AR acknowledges support from the Simons Foundation. We thank A. Alan Middleton, Gilles Tarjus, and Karin A. Dahmen for helpful discussions. 
\end{acknowledgments}
\bibliography{RFIM2D}{}

\clearpage
\appendix

\section{\label{app:invariant} Invariant Scaling Combinations}
\subsection{\label{app:powerlaw} Power Law Form}
As our invariant parameter combinations are unorthodox, we provide here a thorough derivation and a comparison to the usual power law `homogeneous' variables seen at the usual hyperbolic fixed points. The invariant scaling combinations corresponding to traditional power law scaling may be simply derived from the flow equations in 3 and higher dimensions. We have
\begin{equation}
	\begin{split}
		\frac{dw}{d\ell}&=\frac{1}{\nu} w\\
		\frac{ds}{d\ell}&=- \frac{1}{\sigma\nu}s\\
		\frac{dh}{d\ell}&=\frac{\beta\delta}{\nu}h
	\end{split}
\end{equation}
\noindent Taking $(dw/d\ell)/(ds/d\ell)$ and integrating gives
\begin{equation}
	\int_{w_0}^{w^*}\frac{dw}{(1/\nu) w} = \int_{s_0}^{s^*}\frac{ds}{(-1/\sigma\nu) s} 
\end{equation}
\noindent Performing the integral and working through the algebra
\begin{equation}
	\begin{split}
		&\log\ w^*-\log\ w_0 = -\sigma(\log\ s^*-\log\ s_0)\\
		&\Rightarrow  \sigma \log(s_0)+ \log\ w_0= \sigma \log\ s^*+\log\ w^*\\
		&\Rightarrow s_0^\sigma w_0=\textrm{constant}
	\end{split}
\end{equation}
\noindent where $(w^*, s^*)$ corresponds to the fixed point of the RG and is hence a constant. The invariant scaling combination in this instance is thus
\begin{equation}
	s^\sigma w
\end{equation}
which agrees with the results in 3 and higher dimensions~\cite{Perkovic96}. Similarly for $h$ we have
\begin{equation}
	\int_{w_0}^{w^*}\frac{dw}{(1/\nu) w} = \int_{h_0}^{h^*}\frac{dh}{(\beta\delta/\nu) h} 
\end{equation}
\noindent Performing the integral and working through the algebra
\begin{equation}
	\begin{split}
		&\beta\delta(\log\ w^*-\log\ w_0) = \log\ h^*-\log\ h_0\\
		&\Rightarrow  \log\ h_0 - \beta\delta \log\ w_0= \log\ h^*- \beta\delta \log\ w^*\\
		&\Rightarrow h_0 w_0^{-\beta\delta}=\textrm{constant}
	\end{split}
\end{equation}
\noindent The invariant scaling combination is hence
\begin{equation}
	h/ w^{\beta\delta}
\end{equation}
\noindent which again agrees with the literature~\cite{Perkovic96}.
\subsection{\label{app:truncated} Transcritical Form}
The flow equations using the transcritical form for the disorder are as follows
\begin{equation}
	\begin{split}
		&\frac{dw}{d\ell}= w^2+B w^3\\
		&\frac{ds}{d\ell}= -d_f s-C s w\\
		&\frac{dh}{d\ell}= \lambda_h h+F h w
		\end{split}
\end{equation}
As before, we take the integral of $dw/d\ell$ over $ds/d\ell$ and obtain
\begin{equation}
	\int_{s_0}^{s^*}(1/s)\ ds = \int_{w_0}^{w^*}\frac{-d_f-Cw}{w^2+ B w^3}\ dw
\end{equation}
\noindent Solving for $s_0$ we have
\begin{equation}
	s_0 = \bigg{(}B+\frac{1}{w_0}\bigg{)}^{-B d_f+C}\exp \bigg{(}\frac{d_f}{w_0}\bigg{)}f(w^*, s^*)
\end{equation} 
\noindent where $f(w^*, s^*)$ denotes a function of $w^*$ and $s^*$ and is therefore constant. The invariant scaling combination in this case is then
\begin{equation}
	\frac{s}{\Sigma_{\textrm{th}}(w)}
\end{equation}
\noindent where
\begin{equation}
	\Sigma_{\textrm{th}}(w) = \bigg{(}B+\frac{1}{w}\bigg{)}^{-B d_f+C}\exp \bigg{(}\frac{d_f}{w}\bigg{)}
\end{equation}
\noindent Likewise for $h$ we obtain an invariant scaling combination
\begin{equation}
	\frac{h}{\eta_{\textrm{th}}(w)}
\end{equation}
\noindent where
\begin{equation}
	\eta_{\textrm{th}}(w) = \bigg{(}B+\frac{1}{w}\bigg{)}^{B \lambda_h-F}\exp \bigg{(}-\frac{\lambda_h}{w}\bigg{)}
\end{equation}
\subsection{\label{app:well-behaved} Alternative Transcritical Form}
Applying our methods to the 2D equilibrium RFIM, we find that the fixed point is given by a pitchfork bifurcation corresponding to
\begin{equation}
\frac{dw}{d\ell}= w^3-D w^5
\end{equation} 
In this instance, however, the behavior of the correlation length suggests an alternative choice for the normal form 
\begin{equation}
\frac{dw}{d\ell}= \frac{w^3}{1+D w^2}
\end{equation}
as discussed in ~\cite{Raju17}. This form, while retaining the pitchfork behavior, produces a well behaved correlation function that is also able to capture higher order corrections to scaling which we expect to become important further from the critical point. We may apply the same procedure in the non-equilibrium case, although the function for the correlation length here appears well behaved. This yields an alternative form for the transcritical bifurcation given by
\begin{equation}
	\begin{split}
		&\frac{dw}{d\ell}= \frac{w^2}{1-B w}\\
		&\frac{ds}{d\ell}= -d_f s-C s w\\
		&\frac{dh}{d\ell}= \lambda_h h+F h w
		\end{split}
\end{equation}
We can integrate the first equation to a final point $\ell^*, w^*$ to
find the divergence of the correlation length $\xi(w_0) = \exp(\ell^*)$:
\begin{equation}
	\int_{w_0}^{w^*} (1/w^2 - B/w)\ dw = \int_{0}^{\ell^*} d\ell
\end{equation}
\begin{equation}
\begin{aligned}
	\ell^* &= (-1/w - B\log w)|_0^{\ell^*}\\
		& = 1/w^0 + B \log w^0 - \mathrm{constant}
\end{aligned}
\end{equation}
\begin{equation}
	\label{eq:AltXi}
	\xi_\textrm{alt} = \exp(\ell^*) \propto w_0^B \exp(1/w_0).
\end{equation}
As before, to determine $\Sigma(w)$, we take the integral of $dw/d\ell$ over $ds/d\ell$ and obtain
\begin{equation}
	\int_{s_0}^{s^*}(1/s)\ ds = \int_{w_0}^{w^*}\frac{-d_f-Cw}{w^2/(1- B w)}\ dw
\end{equation}
\noindent Solving for $s_0$ we have
\begin{equation}
	s_0 =w_0^{B d_f -C} \exp \bigg{(}\frac{d_f}{w_0}+B C w_0\bigg{)}f(w^*, s^*)
\end{equation} 
\noindent where $f(w^*, s^*)$ denotes a function of $w^*$ and $s^*$ and is therefore constant. The invariant scaling combination in this case is then
\begin{equation}
	\frac{s}{\Sigma_{\textrm{alt}}(w)}
\end{equation}
\noindent where
\begin{equation}
	\Sigma_{\textrm{alt}}(w) = w^{B d_f -C}\exp \bigg{(}\frac{d_f}{w}+B C w\bigg{)}
\end{equation}
\noindent Likewise for $h$ we obtain an invariant scaling combination
\begin{equation}
	\frac{h}{\eta_{\textrm{alt}}(w)}
\end{equation}
\noindent where
\begin{equation}
	\eta_{\textrm{alt}}(w) = w^{-B \lambda_h + F}\exp \bigg{(}-\frac{\lambda_h}{w}-B F w\bigg{)}
\end{equation}
\subsection{\label{app:pitchfork} Pitchfork Form}
The flow equations using a pitchfork form for the disorder are as follows
\begin{equation}
	\begin{split}
		&\frac{dw}{d\ell}= w^3+B w^5\\
		&\frac{ds}{d\ell}= -d_f s-C s w\\
		&\frac{dh}{d\ell}= \lambda_h h+F h w
		\end{split}
\end{equation}
As before, we take the integral of $dw/d\ell$ over $ds/d\ell$ and obtain
\begin{equation}
	\int_{s_0}^{s^*}(1/s)\ ds = \int_{w_0}^{w^*}\frac{-d_f-Cw}{w^3+B w^5}\ dw
\end{equation}
\noindent Solving for $s_0$ we have
\begin{equation}
	\begin{split}
		s_0 \sim &w_0^{B d_f}(1+B w_0^2)^{-\frac{B d_f}{2}} \\
		&\times \exp \bigg{(}\frac{d_f}{2w_0^2}+\frac{C}{w_0}+\sqrt{B} C \arctan(\sqrt{B}w_0)\bigg{)}
	\end{split}
\end{equation} 
\noindent The invariant scaling combination in this case is then
\begin{equation}
	\frac{s}{\Sigma_{\textrm{pf}}(w)}
\end{equation}
\noindent where
\begin{equation}
	\begin{split}
		\Sigma_{\textrm{pf}}(w)  = & w^{B d_f}(1+B w^2)^{-\frac{B d_f}{2}} \\
		& \times \exp \bigg{(}\frac{d_f}{2w^2}+\frac{C}{w}+\sqrt{B} C \arctan(\sqrt{B}w)\bigg{)}
	\end{split}
\end{equation}
\noindent Likewise for $h$ we obtain an invariant scaling combination
\begin{equation}
	\frac{h}{\eta_{\textrm{pf}}(w)}
\end{equation}
\noindent where
\begin{equation}
	\begin{split}
		\eta_{\textrm{pf}}(w) = & w^{-B \lambda_h}(1+B w^2)^{\frac{B \lambda_h}{2}} \\
		&\times \exp \bigg{(}-\frac{\lambda_h}{2w^2}-\frac{F}{w}-\sqrt{B} F \arctan(\sqrt{B}w)\bigg{)}
	\end{split}
\end{equation}

\section{Simulations}
\label{app:sims}

Experience simulating the RFIM on a square lattice has revealed a propensity for faceting in which the shape of the avalanche size distribution becomes dependent on properties of the lattice for small avalanche sizes.  To mitigate this effect, we perform simulations on a periodic Voronoi lattice (Fig.~\ref{fig:Voronoi}) where, for each value of $r$, we consider 100 distinct lattices of size 1000x1000.  Voronoi cells were chosen by generating random coordinates between 0 and 1 and constructing the cells with a 2D implementation of Voro++~\cite{Rycroft09} provided by C. H. Rycroft.  Examples of the avalanche behavior for different values of $r$ are shown in Figure \ref{fig:avalanches}. 

We note that much larger simulations have been done on the square lattice, including a thorough analysis of results from a  $131,072^2$ lattice~\cite{Spasojevic11,Spasojevic11-2}. In analysis of in house simulations on a square lattice, however, we encountered long, unnaturally straight avalanche boundaries. We found these distortions strongly affected the shape of the size distribution for small disorders and served to effectively decreased the system size, a difficulty which became dramatically more pronounced as the disorder decreased. In addition to lattice dependent effects infecting the distributions for larger and larger avalanche sizes approaching the critical point, this effective reduction of system size encouraged the use of a Voronoi lattice. 

From the simulations we extract two quantities of interest: the area weighted avalanche size distribution $A(s|r)$~\cite{Chen11} and the change in magnetization of the sample with respect to the field $\frac{dM}{dh}(h|r)$. Alternatively, we may write these as $A(s|w)$ and $\frac{dM}{dh}(h|w)$ where $w$ is a function of $r$ as defined earlier. 

Lattice effects are a major feature in the two-dimensional NE-RFIM. On
the square lattice, the strong faceting effects due to the lattice 
distorted the avalanche size distribution, effectively giving a short-distance
cutoff not of the lattice constant, but of the typical length $\xi_F$ of the 
straight, horizontal or vertical portions of the avalanche boundaries.
On the random Voronoi lattices we simulate, the stochastic bond
configurations introduce a randomness in the connectivity 
of the network, which we argue here may lead to an effective disorder that
does not vanish even as $r_c\to 0$. One could envision off-lattice
simulations or experiments that could bypass these effects. Here, instead,
we shall briefly explore the faceting and intrinsic disorder, and 
speculate about strategies one might use to minimize their effects.

\subsection{\label{app:Faceting} Faceting}

The experimental systems to which we apply our avalanche model typically
do not have an important underlying lattice anisotropy. The length
scales of the domain wall pinning and avalanches are typically much larger
than the atomic scale, and the materials are often amorphous or
polycrystalline. We thus do not want an underlying crystalline lattice 
dominate the behavior on long length scales.

Often there are emergent symmetries at critical points. Lattice models
(Ising, QCD) break rotational symmetry, but the emergent fluctuations
on long length scales restore this symmetry -- the symmetry of the fixed
point is greater than that of the Hamiltonian. (The short-distance asymmetry
is an irrelevant perturbation.) This is not always the
case, as was vividly illustrated by diffusion-limited aggregation:
simulations on a square lattice led to `dust balls' of the form of giant
crosses~\cite{Meakin86,MeakinBRS87}. There the anisotropy effects for
small-scale simulations appeared unimportant; it was only when large
simulations were visualized that the problem became apparent. 

Are lattice effects relevant or irrelevant for our 2D NE-RFIM? In particular, 
simulations show that avalanche boundaries have long, straight segments
in the horizontal and vertical directions. How do the typical lengths of these
segments $\xi_F$ compare to the avalanche correlation length $\xi$ as we
approach the critical point $r_c$? Can we modify the details of the model
to minimize the effects of this faceting?

We can estimate the length scale $\xi_F$ as a function of disorder on the square
lattice following Drossel and Dahmen~\cite{Drossel98,BlosseyKD98}.
Consider a distribution $p(h|r)$ of random fields parameterized by disorder
$r$. (For most simulations this is taken to be a normal distribution
$p(h|r) = (1/\sqrt{2 \pi r}) \exp(-h^2/2r^2)$.)
On a square lattice, a flat initial horizontal or vertical interface can 
nucleate a pair of steps by flipping a spin at its edge; such a spin has
only one neighbor up, so it must have a random field large enough
that $h-2J+H > 0$. The density of these nucleating sites
at an external field $H$ is thus $\rho_1(H,r) = \int_{2J-H}^\infty p(h) dh$.
Once nucleated, the steps can grow outward until they reach a pinning spin that
will not flip until three of its neighbors are up. A spin will not flip with
two neighbors up if $h<-H$, so these pinning sites happen with density
$\rho_3(H,r) = \int_{-\infty}^{-H} p(h) dh$. By considering the possible
two-dimensional static fronts that avoid the nucleating sites and `turn right'
only on the pinning sites~\cite{Drossel98}, Drossel and Dahmen argue
that for small disorder the interface must depin when 
$\rho_3 = \rho_1^2 / (1-\rho_1)$. They note that the length of the straight
interface segments goes as $\xi_F \sim \rho_1^{-1}$.

If we use the traditional normal distribution, so 
$\rho_1(H,r) \approx (r/\sqrt{2\pi (2J-H)^2}) \exp(-(H-2J)^2/2r^2)$
and 
$\rho_3(H,r) \approx (r/\sqrt{2\pi H^2}) \exp(-H^2/2r^2)$
this implies 
$H_c \to_{r\to0} 2 (2-\sqrt{2}) J \approx 1.172 J$, and $(H_c-2J)$. 
This tells us that the facet length scale for a normal distribution is
\begin{equation}
\begin{aligned}
\xi_F \sim & r \exp((H_c-2J)^2/2r^2) = r \exp(2 (3-2\sqrt{2}) J^2/2r^2) \\
      & \approx r \exp(0.343 J^2/2r^2).
\end{aligned}
\end{equation}
Under the hypothesis that $r_c=0$, this suggests that faceting is a relevant
perturbation, diverging faster as $r\to 0$ than our predicted correlation length
divergence $\xi \sim (r/r_s)^B \exp(r_s/r)$ (Eq.~\ref{eq:AltXi}). This
is consistent with simulations which macroscopically show rough avalanche
boundaries: just as for diffusion limited aggregation, the lattice anisotropy
may be small far from the critical point and become dominant only for
large systems close to critical. (An effect that grows faster as one
approaches the critical point also dies faster as one departs from
the critical region.) Presumably all the 
the avalanches would eventually become nearly rectangular at
sufficiently large lattice sizes and low disorders. Even if $r_c$ is not
zero, it is definitely small for the square lattice, so $\xi_F$ is large.
Since scaling is not expected on lengths smaller than $\xi_F$, the effective
simulation size is reduced by a factor of $\xi_F(r_c)^2$ (the facet length
replacing the lattice cutoff), making it valuable to measure and reduce it.

It would be interesting to measure the distribution of segment lengths for
avalanches in the 2D square-lattice NE-RFIM to test these predictions. One
could also choose random field distributions $p(r)$ that have fatter tails.
If $p(r) \sim \exp(-|H|/r)$ for large $|H|$, the same argument predicts
$\xi_F \sim \exp(2J/3r)$. Comparing to the expected avalanche size 
divergence $\xi \sim w^B \exp(1/w) = (r/r_s) \exp(r_s/r)$ (Eq.~\ref{eq:AltXi}),
and again assuming $r_c=0$, the facet length could diverge 
more slowly than the avalanche size if the nonuniversal scale factor
$r_s > 2J/3$. Using a Cauchy distribution $p[h] = (2 r/\pi) / (r^2 + h^2)$
with very fat tails would yield $\xi_F \sim \sqrt{J/r}$, a much weaker
divergence (smaller facets). One could also generalize Drossel and 
Dahmen's results to other ordered lattices, to examine whether they 
are less susceptible to faceting.

One should be warned that many of these ideas were explored by
Matthew Kuntz in his 1999 Ph.D.\ 
thesis~\cite{KuntzPhD}. He did not measure $\xi_F$, but he
did explore the behavior of the avalanche size distribution, both for
different lattice structures and for a Cauchy distribution of random fields,
with simulations up to size $45000^2$. He found that the disorder-dependent
shape of the avalanche size distribution (the growing bump that prevented
collapses like those in Fig.~1 of the main text) was remarkably invariant
to lattice structure or disorder. Also, the intra-avalanche correlation 
function along the axes equalled the correlations along diagonals after only
a few lattice spacings. Future work, thus, may uncover other explanations
for the striking differences in behavior of the 2D NE-RFIM on regular and
random lattices.

\subsection{\label{app:Disorder} Intrinsic disorder from the Voronoi lattice}

\begin{figure}
	\includegraphics[width=0.4\textwidth]{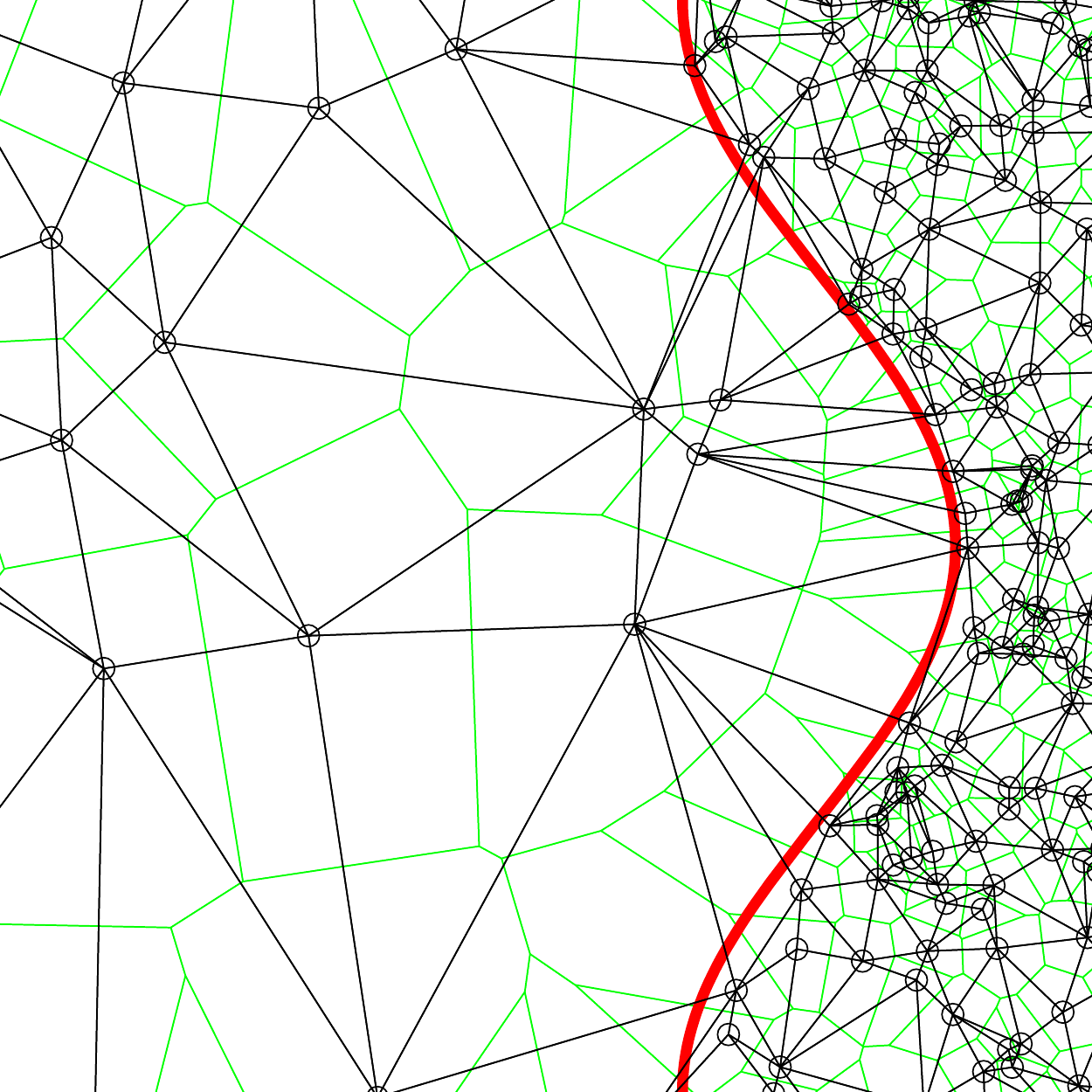}
	\caption{{\bf Density fluctuations of Voronoi cells give effective
	random fields.} Our model puts spins on a randomly chosen lattice
	of sites (circled), and assigns bonds to their Voronoi neighbors
	(black lines connecting sites). The energy per unit length of a
	domain wall between up and down spins will be proportional to
	the square root of the density of points. Here the red curve 
	denotes a boundary between
	artificially created low- and high-density regions.
	An invading avalanche of `up' spins entering from the left will
	likely get pinned as it approaches the red curved boundary,
	just as if there were random fields pointing downward along that curve.}
	\label{fig:densityJumpVoronoi}
\end{figure}

While our fits are consistent with $r_c=0$, Fig.~4 in the main text
is strongly suggestive of a simple scaling with $r_c<0$. This is,
of course, precisely what one would expect if the lower critical dimension
is greater than two: at no disorder would an infinite system show 
system-spanning avalanches, and even in the limit of zero disorder we
would not find broad distribution of avalanches sizes and scaling. 
This is thus a serious concern.

As we have
focused on simulations all of the same size ($10^6$ spins), it is of
course possible that the apparent $r_c<0$ is a (surprisingly large)
finite size effect, and would go away for larger systems. We suspect
that this is not the case: instead, the negative disorder is due to our
choice of a disordered bond network in the quest to remove faceting effects. 

Even in the absence of random fields, our Voronoi lattice simulations 
have disorder in the bond connectivity. On the left of
Fig.~\ref{fig:densityJumpVoronoi}, spins on different sites have different
numbers of neighbors. At low disorder, a spin with eight neighbors will
need at least four of the neighbors to flip, while a spin with four neighbors
would need only two. For the equilibrium model, bond disorder is in a
different universality class than random field disorder, because bond
disorder in zero field preserves the up-down symmetry while random field
disorder breaks that symmetry for each individual system, preserving it
only for the ensemble. The NE-RFIM, however, has an external field growing
from $H=-\infty$; this external field history breaks the up-down symmetry,
and indeed (as discussed above) the critical field is positive, not zero.
Simulations~\cite{Vives95}
have shown, indeed, that the random-bond Ising model is in 
the same universality class as the random-field Ising model~\cite{Dahmen96}.

Fig.~4 in the main text suggests that the intrinsic randomness of our
Voronoi lattice shifts $r_c$ from zero to roughly $-J/2$ -- corresponding
to mean fluctuations of half of a bond.
A spin at a growing front needs an external field
to counteract the imbalance in the number of up and down neighbors; a spin
with more neighbors will typically have a larger imbalance. 
The root-mean-square
fluctuations in the number of bonds connecting a spin to its neighbors in
a 2D Voronoi lattice with random sites is $1.334$~\cite{Brakke}, making
the contribution of $-J/2$ in the effective random field entirely plausible.
As mentioned
in the text, one could test whether bond coordination randomness
causes avalanches to remain finite in size by using
fairly feasible simulations of 
$\sim 4\, 10^8$ spins, if the challenge of building such a large Voronoi
lattice could be surmounted.  

Disordered lattices have also been explored for the NE-RFIM by Kurbah,
Thongjaomayum and Shukla~\cite{Kurbah15}. Site dilution of one
sublattice of the triangular lattice allows them to explore 
average coordinations $Z$ continuously varying between three and six. As found
in early work on the NE-RFIM on the Bethe lattice~\cite{DharSS97}, they
find a critical disorder at $Z_c=4$,
with no transition for $Z=3$ and fitting to power-law scaling for $Z \ge 4$,
with results consistent with the same critical exponents for all 
coordinations showing a transition. For $Z=3$ and $Z=6$, Kurbah {\em et al.}
are subject to the same faceting problems concerns on the square
lattice (and the triangular lattice~\cite{KuntzPhD}). For intermediate
values, they have both rotational anisotropy and disorder. Later work
by Shukla and Thongjaomayum~\cite{Shukla16} study the NE-RFIM on
the diluted Bethe lattice, and find $Z_c=3$ -- lower than $Z_c=4$ for the
diluted triangular lattice, making the role of coordination in governing
the behavior suspect. We would expect that the systems studied by Kurbah
{\em et al.} will exhibit a negative critical random field strength $r_c<0$
for all $3<Z<6$ -- zero critical {\em disorder} but with contributions
to the disorder both from the random field and from the random bonds.

Consider, for example, density fluctuations (Fig.~\ref{fig:densityJumpVoronoi}).
The sites far to the left and far to the right of the jump
in density (red curve)
have on average six neighbors (by Euler's theorem), and thus have the same
critical field at which the interface depins. However, the sites 
just to the left of the jump have typically more than six neighbors, and
thus will demand a higher critical field in order to flip -- just as if
they had random fields pointing downward. It could be useful to explore
Voronoi simulations whose lattices have been tailored to have reduced
density fluctuations. These could be generated by adding Gaussian noise
to regular lattices, or perhaps by generating lattices from jamming 
simulations. One could also {\em increase} the disorder by artificially 
correlating the positions of the spins, to see if $r_c$ becomes more negative
-- perhaps allowing simulations with $r=0$ to be realized for relatively
small lattices. 

How can we argue that $r_c<0$ is due to intrinsic disorder, and not partly
also due to the lower critical dimension being higher than two? Our primary
argument is the excellent (albeit non-power-law) scaling over a decade in 
disorder (Figs.~1-3 in the main text), and the excellent collapses found
under the presumption that the RG flows have a lower critical dimension 
(transcritical bifurcation) in $d_u=2$.

\end{document}